\documentclass[journal = jpccck, manuscript=article]{achemso}

\usepackage{graphicx}
\usepackage{color}
\usepackage{epstopdf}
\DeclareGraphicsExtensions{.png}

\title[Hydrophobic Ice Confined between Graphene and MoS$_{2}$]{Hydrophobic Ice Confined between Graphene and MoS$_{2}$}

\author{Pantelis Bampoulis}
 \email{p.bampoulis@utwente.nl}
\affiliation{ 
Physics of Interfaces and Nanomaterials, MESA+ Institute for Nanotechnology, University of Twente,  P.O.Box 217, 7500AE Enschede, The Netherlands.
}%
 \alsoaffiliation{Physics of Fluids and J.M. Burgers Centre for Fluid Mechanics, MESA+ Institute for Nanotechnology, University of Twente,  P.O.Box 217, 7500AE Enschede, The Netherlands.}

\author{Vincent J. Teernstra}
\affiliation{ 
Physics of Interfaces and Nanomaterials, MESA+ Institute for Nanotechnology, University of Twente,  P.O.Box 217, 7500AE Enschede, The Netherlands.
}%

\author{Detlef Lohse}
\affiliation{%
Physics of Fluids and J.M. Burgers Centre for Fluid Mechanics, MESA+ Institute for Nanotechnology, University of Twente,  P.O.Box 217, 7500AE Enschede, The Netherlands.
}%
\alsoaffiliation{Max Planck Institute for Dynamics and Self-Organization, Am Fassberg, 37077 G{\"o}ttingen, Germany}

\author{Harold J. W. Zandvliet}
\affiliation{ 
Physics of Interfaces and Nanomaterials, MESA+ Institute for Nanotechnology, University of Twente,  P.O.Box 217, 7500AE Enschede, The Netherlands.
}%

\author{Bene Poelsema}
\affiliation{ 
Physics of Interfaces and Nanomaterials, MESA+ Institute for Nanotechnology, University of Twente,  P.O.Box 217, 7500AE Enschede, The Netherlands.
}%

\begin{document}

This article may be downloaded for personal use only. Any other use requires prior permission of the author and ACS publications. The following article appeared in . Phys. Chem. C, 2016, 120 (47), pp 27079–27084 and may be found at 

http://pubs.acs.org/doi/abs/10.1021/acs.jpcc.6b09812.

\begin{abstract}
The structure and nature of water confined between hydrophobic molybdenum disulfide (MoS$_2$) and graphene (Gr) are investigated at room temperature by means of atomic force microscopy (AFM). We find the formation of two dimensional (2D) crystalline ice layers. In contrast to the hexagonal ice `bilayers' of bulk ice, these 2D crystalline ice phase consists of two planar hexagonal layers.  Additional water condensation leads to either lateral expansion of the ice layers or to the formation of 3D water droplets on top or at the edges of the two-layer ice, indicating that water does not wet these planar ice films. The results presented here are in line with a recent theory suggesting that water confined between hydrophobic walls forms 2D crystalline two-layer ice with a non-tetrahedral geometry and intrahydrogen bonding. The lack of dangling bonds on either surface of the ice film gives rise to a hydrophobic character. The unusual geometry of these ice films is of great potential importance in biological systems with water in direct contact with hydrophobic surfaces.
\end{abstract}

\section{Introduction}
The properties and structure of water at planar surfaces have been extensively investigated using surface science techniques \cite{Hodgson2009381, Thiel1987211, Verdaguer20061478, Feibelman201034, Berkelaar20151017} due to enormous interest in many fields including environmental sciences, lubrication \cite{Lee201076}, nanofluidics \cite{Prakash2008441} and biology \cite{Chaban20115647}. The structure and nature of water is often affected by the surface's structure and wettability, pressure and temperature. For example, due to the strong binding of water to hydrophilic substrates, the structure of water near the surface is influenced by the substrate. This often results in non-tetrahedral bonding geometries, as has been observed for crystalline hydrophobic ice films grown on Pt(111) at cryogenic temperatures \cite{Kimmel2005}. 

The lack of bonding of water to a hydrophobic substrate as well as stratification induced by the substrate (in its vicinity) can also lead to the growth of ice films with a non-tetrahedral geometry and remarkable properties. Kimmel et al. \cite{Kimmel200912838} observed metastable crystalline ice grown at 100-135 K on Gr/Pt(111). This crystalline ice consists of two flat hexagonal layers of water molecules. The hexagons of each layer are placed directly on top of each other. In this structure each molecule forms three in plane hydrogen bonds with each nearest neighbors and one out of plane hydrogen bond with the molecule in the opposite layer. This way the number of hydrogen bonds is maximized. There are no dangling bonds on either surface of the ice. The ice is considered hydrophobic since adding a third water layer is energetically not favorable. Water molecules that are located on top of the double ice layer will either diffuse towards the edge of the ice and incorporate into the double layer or will form 3D clusters. Similar ice films have also been observed on a Au(111) substrate \cite{Stacchiola200915102} at cryogenic temperatures and it has been hypothesized to exist on any hydrophobic planar surface due to the weak (van der Waals) interactions of the water molecules with the atoms of the surface \cite{Kimmel200912838}. Molecular dynamics predicted the existence of these ice films in confined hydrophobic geometries \cite{Koga19975262} that can be stable at room temperature \cite{Kimmel200912838, Giovambattista2009}, but they have yet to be experimentally observed. 

It is well known that the properties of confined water are in many cases dramatically different from their bulk counterpart. These properties include complex phase behavior, non-tetrahedral bonding geometry, novel phase transitions and anomalous self diffusion \cite{Zhao20142505, Cambré2010, Mukherjee2007, Giovambattista2009, Koga2000564, Cicero20081871}. Even though most knowledge comes from MD simulations and DFT calculations, recent experimental work of confined water in carbon nanotubes and Gr (slightly hydrophobic with bulk contact angle ranging from 70$^o$ to 90$^o$) (and other 2D materials) has provided information on the state of water confined between hydrophilic/hydrophobic \cite{Bampoulis2015, Severin2012774, Bampoulis2016, Xu20101188, Bampoulis20166762, Dollekamp20166582} and hydrophobic/hydrophobic \cite{Cao20115581, Algara-Siller2015443} interfaces. For instance, Algara-Siller et al.\cite{Algara-Siller2015443} provided evidence that water confined between two graphene surfaces acquires a square lattice, attributed to the large exerted pressure induced by the graphene surfaces. Despite significant efforts a systematic understanding of the influence of confinement on this rich behavior is still lacking. Understanding the vast amount of ice phases and the dynamics of ice structures is crucial for many fields including life sciences, environmental sciences, condensed matter physics, lubrication and nanofluidics. For example, ice layers on surfaces are used for the preparation of supported metal \cite{Palmer20082278} and oxide nanoparticles \cite{Kaya20075337}. Moreover, water diffusion in hydrophobic nanopores is of great relevance for biological systems and nanofluidics \cite{Chaban20115647}. In the case of 2D materials, water adsorbates or intercalates can influence their electronic properties. For example, ice on Gr or under the Gr can electronically dope the Gr \cite{Goncher20131386, Bollmann20153020}. The type of doping and level depends crucially on the structure and state of the ice.

In this paper we provide compelling experimental evidence that water confined at room temperature between Gr/MoS$_2$ forms a unique two-layer (2L) structure with low surface energy due to intrahydrogen bonding inside the ice. Our findings are in line with recent theoretical works, which have predicted the existence of a hydrophobic double layer ice with non-tetrahedral geometry in a hydrophobic confinement and at room temperature \cite{Kimmel200912838, Giovambattista2009}. Our work paves the way towards an understanding of the complex behavior of confined water structures at RT.

\section{Methods}
Exfoliated Gr flakes (cleaved from highly oriented pyrolytic graphite (HOPG ZYA grade, MikroMasch), consisting of one or a few layers  are deposited on top of a freshly cleaved synthesized 2H-MoS$_2$ (2D Semiconductors) substrate. Like Gr, MoS$_2$ is considered to be hydrophobic. The reported contact angles, for water droplets on this surface, are in the range of 70-90$^o$ \cite{Gaur20144314}. Optical Microscopy (DM2500H materials microscope, Leica, Germany) was used to locate the Gr flakes. Optical Microscopy and tapping mode atomic force microscopy, AFM, (Agilent 5100 atomic force microscope, Agilent) was used to determine the exact number of Gr layers (see supporting information 1). AFM imaging on the Gr/MoS$_2$ system was performed using Nanosensors SSS-FMR-10 with a nominal spring-constant of 2.8 N/m and resonance frequency of 75 kHz. Water droplets and layers on top of the Gr or the MoS$_2$ substrates require softer cantilevers in order to be properly imaged without affecting their shape and size \cite{Lohse2015}. In these experiments we have used MikroMasch HQ:NSC36/AL BS C cantilevers with a nominal spring-constant of 0.6 N/m and a resonance frequency of 65 kHz. The AFM was operated in intermittent contact amplitude modulated mode. The cantilever was excited with a frequency slightly above its resonance frequency. When the tip is in the proximity of the surface, the oscillation amplitude is reduced as compared to the free amplitude, $A_{Free}$, owing to the force field induced by the sample. Scanning of the surface's topography is achieved by keeping a constant setpoint amplitude ($A_{SP}$). For the case of droplets and water layers on the air/substrate interface (without the graphene cover) $A_{Free}$ was kept below 10 nm and $A_{SP}$ was set to $0.95\times A_{Free}$. This way a potential capture of the tip from the interfacial water structures was avoided. Furthermore, AFM imaging was performed inside an environmental chamber. The relative humidity (RH) inside the glove box and the AFM environmental chamber was controlled by an adjustable $N_2$ flow which either bubbled through a water bottle or directly purged into the chamber and was measured using a humidity sensor (SENSIRION EK-H4 SHTXX, Humidity Sensors, Eval Kit, SENSIRION, Switzerland), which has an accuracy of 1.8 \% between 10-90 \% RH. We emphasize that the experiments have been performed at RT.

\section{Results and Discussion}

Figure 1a describes the experimental process. MoS$_2$ and thin Gr flakes were initially exposed to high (80-90\%) RH for a short period of time (5-10 mins) in a home built glove box. Water adsorbs on both surfaces, see the supporting information 2 for details. The Gr flakes are then deposited on top of the MoS$_2$ substrate. The adsorbed water is confined between the two surfaces and the system is ready to be imaged by AFM, as can be seen in figure 1c (we note here that samples made at low (1-30\%) RH do not show any water structures at the interface, see figure 1b). In contrast to water structures on the free graphite or MoS$_2$ substrates (see supp info 2), the confined water films are stable with well defined faceted edges, suggesting that they are ordered, i.e. crystalline. AFM phase images (see inset of figure 1c) suggest that these water structures are indeed between Gr and MoS$_2$. In general, zero to little phase contrast ($\sim 0-1^o$) is observed, which indicates that the probing tip always interacts with the same surface, i.e. Gr. 

\begin{figure}[htb]
\centering
\includegraphics[width=\textwidth]{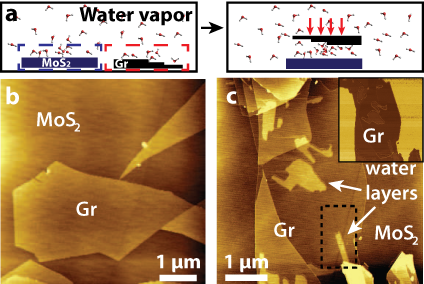}
\caption{(a) Cartoon of the experimental procedure. The Gr and the MoS$_2$ surfaces are exposed to a high RH environment (about 80\%) for a short amount of time (5-10 minutes). Water adsorbs on both surfaces leading to the formation of water layers and droplets. When the two surfaces are brought together these water structures are confined between the two materials. (b) Monolayer Gr on MoS$_2$ deposited under 30\% RH. (c) Monolayer and few layers Gr on MoS$_2$ deposited after the two surfaces were exposed for 5 minutes to high (80-90\%) RH. Water layers are confined between the two surfaces. The inset is the corresponding phase image and it reveals little phase contrast between the water layers and the Gr surface but huge contrast between Gr and MoS$_2$.}
\label{fig:figure1}
\end{figure}

\begin{figure}[htb]
\centering
\includegraphics[width=\columnwidth]{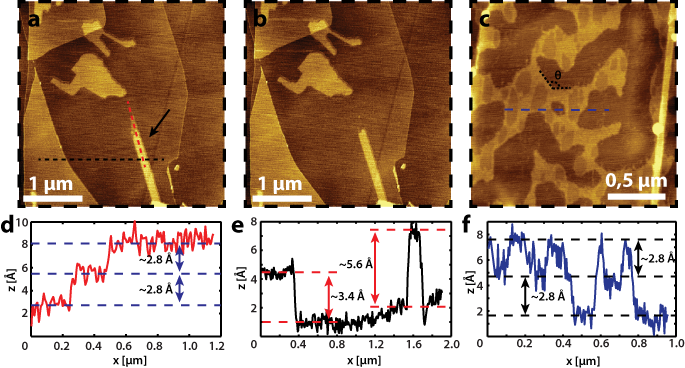}
\caption{(a) and (b) are snapshots during the growth of a water layer indicated with the black arrow. (c) Water layers under few layers of Gr. (d) and (e) and (f) are line profiles taken across the structures of panel (a) (red and black, respectively) and panel (c) (blue) revealing that these structures consist of two layers of water. }
\label{fig:figure2}
\end{figure}

Occasionally, we observe that some of the water structures slowly grow over time, see figures 2a and 2b. We often observe that a propagating front -of a lower height contrast with respect to the initial structure- leads the growth, figure 2a. A movie of the growing ice structure is shown in the supporting information 3. The very rich growth and melting dynamics of these water structures will be discussed in detail in future work. Two height levels can also be seen at various other images and at different areas and samples, for instance, figure 2c shows islands under a few layers of Gr with two distinct height levels. 
\begin{figure}[htb]
\centering
\includegraphics[width=7 cm]{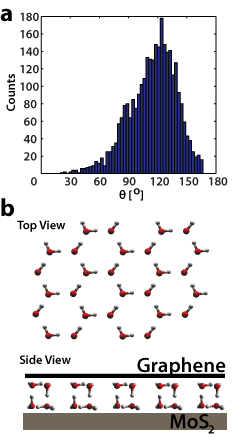}
\caption{(a) Histogram of the distribution of the angles measured between two adjacent edges of the water layers, as indicated in figure 2c. The histogram shows a predominant peak at 120$^o$. (b) Top and side view of the structural model of the 2L ice. The red and white spheres represent O-atoms and H-atoms, respectively.}
\label{fig:figure3}
\end{figure}

Cross-sections taken across the water films, figure 2d, reveal a total height difference with respect to their surroundings of 0.56 nm $\pm$ 0.02 nm (black curve). The cross-sections also reveal that each level (layer) has a height of about 0.28 nm (red and blue curves). These values do not correspond at all to the height of the bilayer of hexagonal ice `Ih-ice' (0.37 nm) or any multiples of it. However, they are rather close to the size of planar ice films, 0.285 nm. Furthermore, they clearly have a hexagonal symmetry as can be seen in the histogram of figure 3a, where a distribution of the angles between two adjacent island edges is shown. A predominant peak at around 120$^o$  is clearly observed, suggesting a hexagonal symmetry. Furthermore, shoulders at approximately 90$^o$ and 60$^o$ are observed. Often, an elongation of the islands is observed, similar to what we saw for water condensation on the bare MoS$_2$ surface, see supporting info 2, indicating that the growth of the islands is kinetically limited.
\begin{figure}[htb]
\centering
\includegraphics[width=\columnwidth]{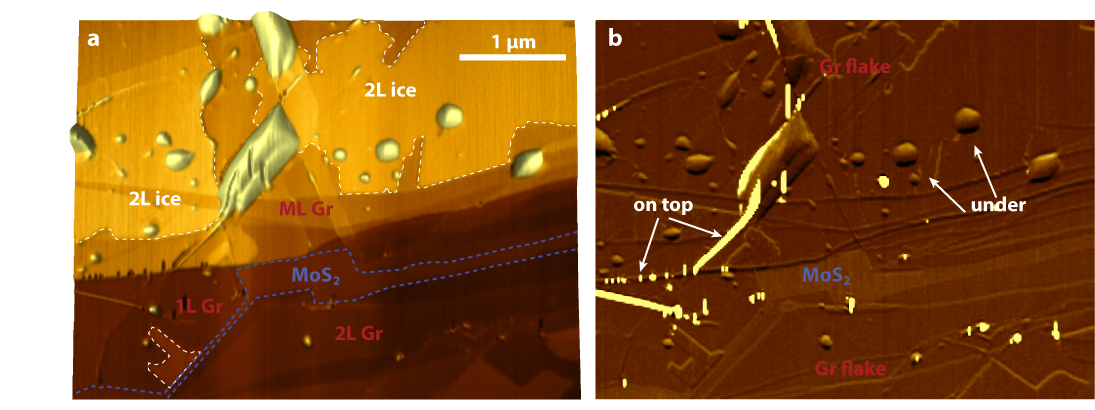}
\caption{(a) A large scale AFM image of a Gr flake with various thickness on MoS$_2$. 2L ice marked with the white dashed lines are under the Gr flake. Occasionally various sizes of water droplets are on top and at the edges of the 2L ice. Droplets between Gr and MoS$_2$ are also observed. (b) The corresponding phase image, both the 2L ice and droplets reveal no apparent phase contrast with respect to the Gr surface. For comparison, water droplets on the MoS$_2$ and Gr reveal a strong phase contrast.}
\label{fig:figure4}
\end{figure}

Our results clearly demonstrate the existence of planar two-layer (2L) crystalline ice films with a hexagonal symmetry in a hydrophobic confinement and at RT. The observed ice structures are reminiscent of a novel ice phase, which has been first observed at low temperature experiments on hydrophobic substrates, such as graphene \cite{Kimmel200912838, Stacchiola200915102}. As has been described first elsewhere \cite{Koga19975262, Kimmel200912838}, these ice films consist of two hexagonal planar layers in registry with each other and with a non-tetrahedral geometry. Every water molecule of the 2L ice forms four hydrogen bonds with its neighbors, three in the same plane and one out of plane with the water molecule of the opposite layer, leading to a self-closed hydrogen bonding network with low surface energy due to the maximized number of hydrogen bonds\cite{Kimmel200912838}. The lattice constant of the 2L ice is 0.48 nm \cite{Kimmel200912838}, in contrast to 0.45 nm of the bilayer Ih. A cartoon of the top and the side view of the 2L ice structure are shown in figure 3b. 

The total area covered by water depends on the exposure time and the distance of the sample from the inlet of the water vapor into the glove box. Longer exposure and at a closer proximity to the inlet results in a larger lateral expansion of the two dimensional ice layers and often to a higher density of water droplets, see figure 4a. Figure 4b shows that all the 2L ice structures and the water droplets are under the Gr, no phase contrast is observed. Water droplets on top of the Gr show strong phase contrast ($\sim 20^o$) and can be dragged by the AFM tip. We have never observed the growth of a third ice layer, which indicates that two layer thick ice is energetically more favorable than three layers thick ice. Occasionally 3D droplets of various sizes are observed on top or at the edges of the 2L ice, figure 4. DFT calculations \cite{Stacchiola200915102} demonstrate that 1L or 3L thick ice are energetically less favored when compared to 2L thick ice: 3L ice is less stable than 2L ice by not less than 0.1 eV/$H_{2}O$ \cite{Stacchiola200915102}, although it is slightly more stable than 1L ice. This last finding contradicts our observations since we have never observed any 3L ice structures, while in contrast, we have seen in many occasions  1L ice structures. We attribute this difference to the fine details of the hydrophobic confinement. The DFT calculations have been performed for ice structures on a hydrophobic solid-vacuum interface, however in our case the second hydrophobic wall might induce another limitation making the 3 layer ice energetically less unfavorable than the 1L ice. The surface free energy of the 2L ice is low. The 2L ice films are only weakly interacting with the substrate or the Gr cover. Owing to these weak interactions, small 2L ice islands are very mobile, see supporting information 3.

\begin{figure}[htb]
\centering
\includegraphics[width=\columnwidth]{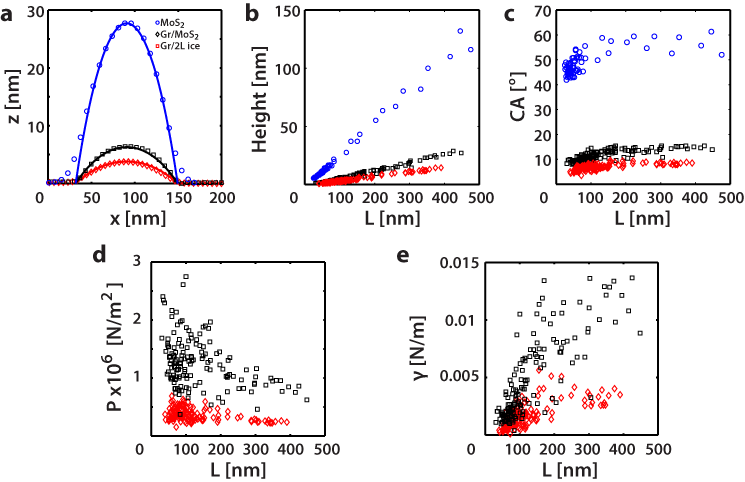}
\caption{(a) Cross sections of droplets between Gr/MoS$_2$ (black), Gr/2L ice (red) and on the bare MoS$_2$ surface (blue). The solid lines are spherical cap fits. (b) Height (H) vs. footprint diameter (L) and (c) contact angle (CA) vs. footprint diameter (L) for all three cases. (d) Pressure (P) vs.  footprint diameter (L) and (e) Adhesion energy ($\gamma$) vs.  footprint diameter (L) for droplets between Gr/MoS$_2$ (black) and Gr/2L ice (red).}
\label{fig:figure5}
\end{figure}
The lack of OH dangling bonds on either of the two surfaces of the ice, gives the 2L ice a hydrophobic character. Water molecules that arrive on top of the two layer ice are weakly bound and will prefer either to diffuse towards the edges where they become part of the ice structure or form a 3D non wetting structure \cite{Kimmel200912838, Kimmel2005}. This is also consistent with our system, since there is no observation of a layer by layer growth but rather a lateral expansion and/or the observation of 3D droplets on top or at the edges of the 2L ice. To test whether the droplets are indeed on top of the ice layers and not surrounded by ice, we have measured the droplet height as a function of footprint diameter and compared it to sessile water droplets on the  MoS$_2$ surface, figures 5a and b. Note here that large droplets under the graphene display a polygonal shape, induced by the graphene cover. In order to avoid any misinterpretation of our results we have restricted our analysis to droplets with a spherical shape, see figure 5a. A clear difference is observed in figures 5a, 5b and 5c, suggesting that the droplets are in between different surfaces, i.e. Gr/MoS$_2$ (black points in figure 5) and Gr/2L ice (red points in figure 5). It is also evident from figure 5c that the contact angle of a water droplet between Gr and MoS$_2$ (black points) is substantially lower than the contact angle of large sessile water droplets on MoS$_2$ (blue points)\cite{Gaur20144314}. In order to check this in more detail we have measured water droplets on the bare MoS$_2$ surface and compared their sizes and contact angles with the confined droplets. We measured contact angles in the range of 60$^o$, i.e., very close but slightly smaller than the reported values for large sessile water droplets. Note here that convolution effects have been taken into account according to ref. \cite{CanetFerrer2014}. We note that especially on the free sessile droplets, different tip-surface interactions can overestimate or underestimate up to some extent these measurements. The large difference in contact angles between sessile droplets on the MoS$_2$ surface and droplets confined between MoS$_2$ and Gr is attributed to confinement effects. The pressure induced by the confinement is substantial\cite{Algara-Siller2015443, Khestanova2016} and therefore capable of deforming small droplets, resulting in  the reduction of their contact angle.

The diameter and height of droplets confined between a 2D membrane, such as Gr, and a supporting substrate depend on the number of molecules, the adhesion energy between the membrane and the substrate, in that case either MoS$_2$ or 2L ice, and the elastic properties of the membrane. The ratio of the height vs the diameter is determined by the balance between van der Waals interactions and the elastic energy of graphene \cite{Khestanova2016}. It has been proposed that this ratio is independent of the substance of the bubble/droplet \cite{Khestanova2016}. Membrane or non-linear plate theories are typically used to determine the adhesion energy of Gr with its substrate, by the use of a trapped bubble or blister \cite{Yue2012, Temmen2014}. Yue et al. \cite{Yue2012} demonstrated that for monolayer Gr covering relative big bubbles the membrane analysis is sufficient. This analysis uses a spherical cap approximation but does not take into account the small but finite bending stiffness of Gr. Therefore it does not describe the bubble's edges accurately but nevertheless provides a simple solution. Using membrane analysis the pressure inside the bubble is obtained by \cite{Yue2012},
\begin{equation} 
P=\frac{E_{2D}H^{3}}{\Phi L^4}+P_0,
\end{equation}
where $E_{2D}$ is the 2D Young's modulus of single layer Gr \cite{Lee2008385}, H and L the height and footprint diameter of the droplet, $P_0$ is the external pressure, $\Phi=\frac{75(1-\nu^{2})}{8(23+18\nu-3\nu^2)}$ and $\nu$ is Poisson's ratio equal to 0.16 for monolayer Gr. We have estimated the pressure inside the droplets between monolayer Gr and MoS$_2$ and monolayer Gr and 2L ice using equation (1), see figure 5d. In both cases smaller droplets display larger pressures, the pressure decays fast as the size increases and saturates at approximately $1\times10^6$ $N/m^2$ for Gr/droplet/MoS$_2$  and $3\times10^5$ $N/m^2$ for Gr/droplet/2L ice.

Using the ideal gas law, the Gr-substrate adhesion energy is directly related to the equilibrium droplet size, since it is a result of the potential energy of the droplet balanced by the interfacial energy. The adhesion energy ($\gamma$) is derived \cite{Yue2012} to be equal to:
\begin{equation} 
\gamma=\frac{5E_{2D}H^{4}}{8\Phi L^4}.
\end{equation}
 The Gr substrate adhesion energy is shown in figure 5e for both cases as calculated from equation (2). It can be seen that Gr adheres better on the MoS$_2$ substrate with an adhesion energy in the order of $\sim$ 0.01 N/m whereas the Gr/2L ice adhesion energy is in the range of $\sim$ 0.003 N/m.
 
\section{Conclusions}
Our scanning probe investigation of confined water is consistent with the existence of a non-tetrahedral ice phase at RT in a hydrophobic confinement. This ice phase consists of two planar hexagonal ice layers. The water molecules of each layer of the ice are stacked on top of each other and have a self closed hydrogen bonding. This leads to weak interactions with the substrate and with other water molecules on top of the 2L ice. Our results are in line with recent MD simulations of ice films confined between hydrophobic walls and low temperature experiments of ice films on graphene and Au(111) substrates. This unusual geometry of the 2L ice may be of importance in biological samples, where water is in direct contact with hydrophobic surfaces, and may be also relevant for water structures on hydrophobic surfaces and at ambient conditions.

\section{Acknowledgements}
We would like to thank the Nederlandse Organisatie voor Wetenschappelijk Onderzoek (NWO, STW 11431) for financial support.

\bibliography{Bibliography}

\section{Supporting Information}

\section{1. Determination of the Graphene Flake Thickness}
\begin{figure}[htb]
\centering
\includegraphics[width=\textwidth]{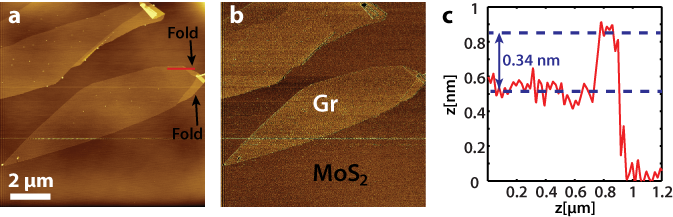}
\caption{(a) Graphene flakes on the MoS$_2$ surface. The graphene flake is folded at one of each sides. (b) The corresponding phase image of (a) revealing two distinct contrasts that correspond to the graphene and MoS$_2$ regions. (c) Line profile of the graphene fold noted with the red solid line in panel (a). }
\label{fig:figure1S}
\end{figure}
In order to determine the thickness of the graphene flake on MoS$_2$ we use AFM images. In principle, AFM cannot provide accurate height interpretation between two different materials, i.e. Gr and MoS$_2$, due to the different tip-surface interactions between the two surfaces. In order to overcome this problem, we use regions where one of the sides of the graphene flake is folded, see for example figure \ref{fig:figure1S} (a). These folds can be often found on exfoliated samples and are a product of the exfoliation process. They can be also manually created by scanning the edge of the graphene flake in contact mode. A line profile measurement across the fold, figure \ref{fig:figure1S} (c), can then be used to exactly determine the thickness of the graphene flake, in this case the flake is a monolayer of graphene and thus the fold a bilayer.    

\section{2. Adsorption of Water on Graphene and MoS$_2$}
\begin{figure}[htb]
\centering
\includegraphics[width=\textwidth]{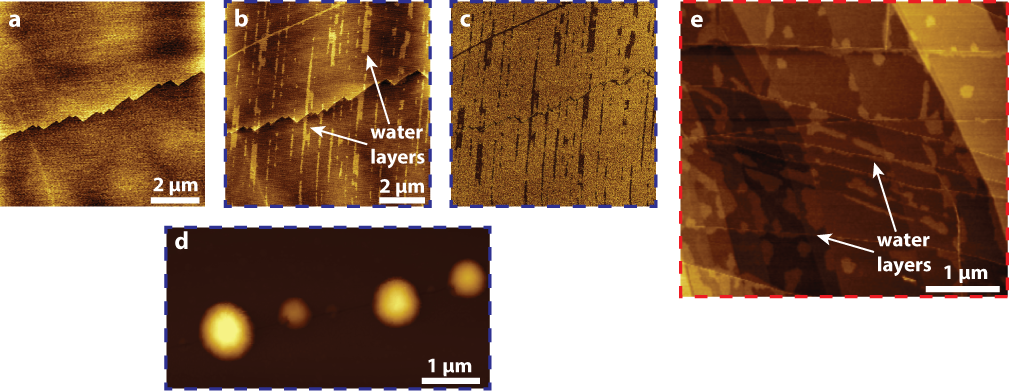}
\caption{(a) AFM topographic image of a freshly cleaved MoS$_2$ measured at ambient conditions. (b) The same location after exposing the surface to 85\% RH for 5 minutes. Elongated water layers are formed along the surface. (c) The corresponding phase image of (b), revealing sharp phase contrast between the water layers and the MoS$_2$ surface. (d) Water droplets formed on the MoS$_2$ surface after long exposure to high RH. (e) AFM image of an HOPG sample after exposure to water vapor for a couple of minutes. Water layers are visible on the whole surface.}
\label{fig:figures2}
\end{figure}

In order to clarify that water adsorbs on both HOPG (or Graphene flakes) and MoS$_2$ surfaces we have exposed them at high relative humidity (80-90 \%) for a couple of minutes and subsequently imaged the surfaces with AFM. The results are shown in figure \ref{fig:figures2}. Water adsorbs in the form of layers or droplets at both surfaces depending on the exposure time but also at the surface's morphology. We would like to note here that in contrast to the well defined ice structures that are confined between Gr and MoS$_2$, these water layers are very dynamic with lousy edges and can be easily disturbed and modified by the probing tip. Note also that they exhibit a large phase contrast with respect to their surroundings and in contrast to the confined structures.

\section{3. Growth of Ice Crystals and Mobility of Small Clusters}
\begin{figure}[htb]
\centering
\includegraphics[width=\textwidth]{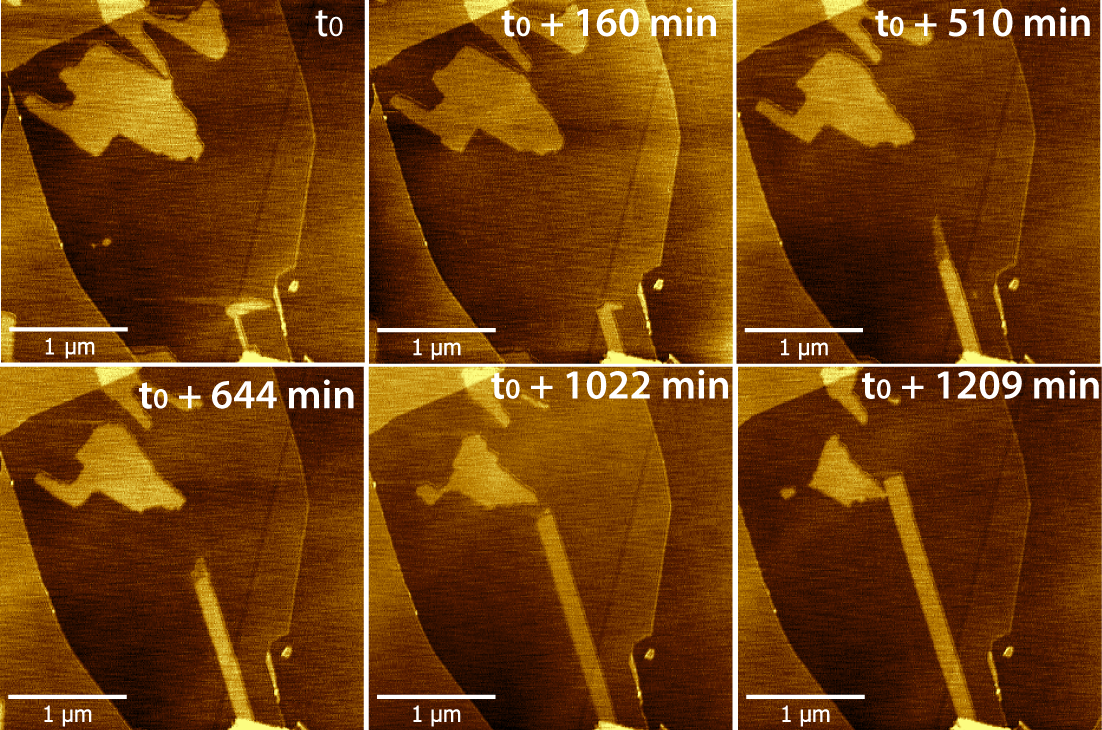}
\caption{(a) AFM image snapshots of the growth of a 2L ice crystal. The 2L ice crystal grows at the expense of surrounding 2L ice crystals.}
\label{fig:figureS3}
\end{figure}
We have recorded the dynamics of the 2L ice crystals by scanning continuously over a long period of time (3 days) with AFM. In figure \ref{fig:figureS3} snapshots of the growth of a 2L ice crystal is shown. The 2L ice crystals grows in one direction (different from the scanning direction by about 30$^o$) acquiring an elongated form. This crystallite appears to grow at the expense of the surrounding crystals, indicating mass transport. As the crystal grows, often, a propagating 1 layer thick front is observed which acts as an apparent precursor of the 2L ice. The complete movie is given as a suppl. movie. Similar dynamics have been observed in  several samples. A detailed study of the growth and melting dynamics of the 2L ice crystals will be part of a future work. 

\begin{figure}[htb]
\centering
\includegraphics[width=\textwidth]{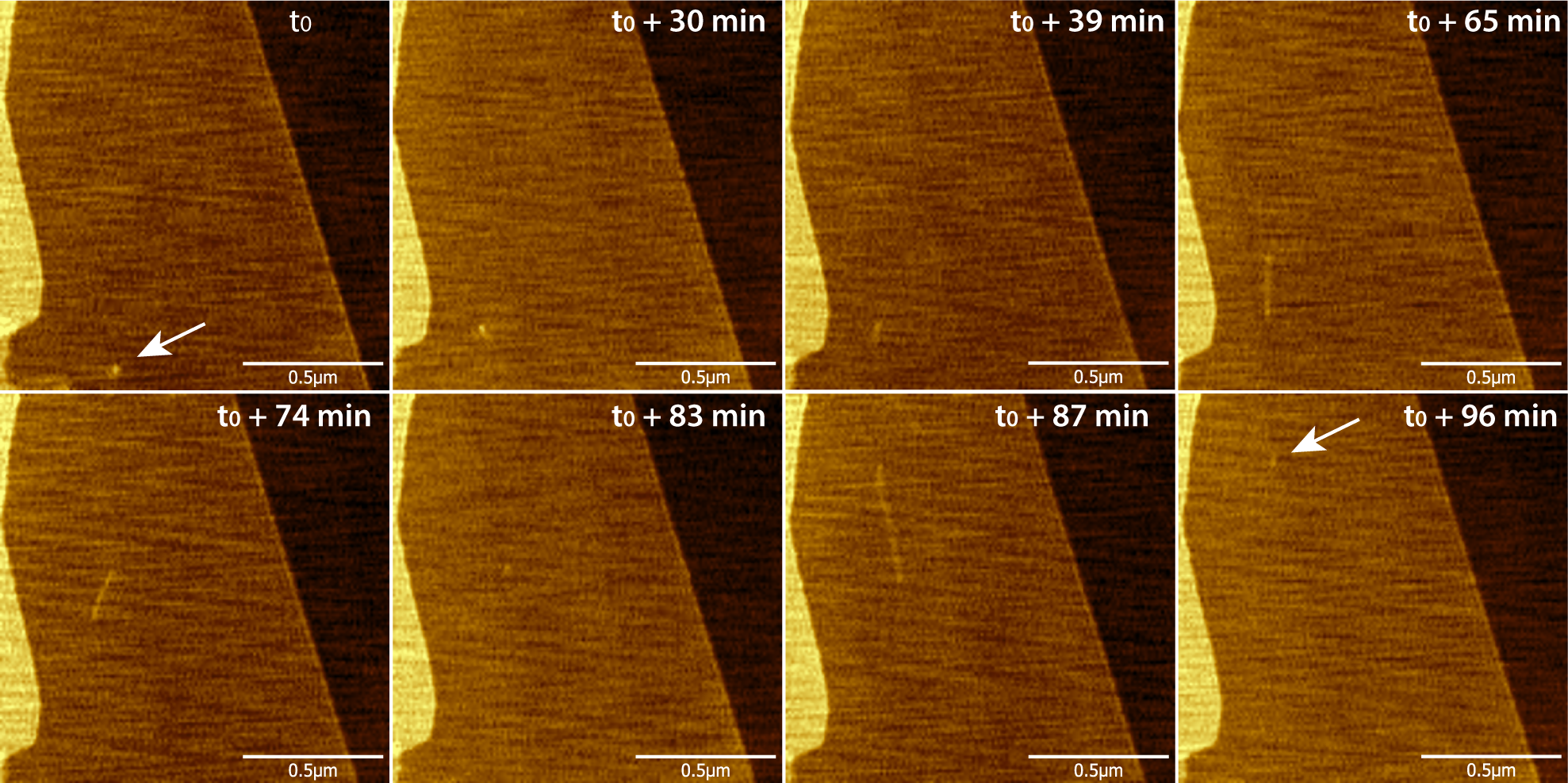}
\caption{(a) AFM image sequence of a small 2L ice crystal. The ice crystal shows rapid movement in between the two surfaces. Eventually it escapes from the interface at the Gr MoS$_2$ boundary.}
\label{fig:figureS4}
\end{figure}

We have also recorded the mobility of small 2L ice crystals. Snapshots are given in figure \ref{fig:figureS4}. The 2L ice crystals moves with steps up to 100s of nanometer within a few minutes of measuring time. The high mobility of small clusters is explained by the weak interaction of the 2L ice with both surfaces (MoS$_2$ and Gr).

\end{document}